\documentclass[nohyperref]{article}

\usepackage{microtype}
\usepackage{graphicx}
\usepackage{subfigure}
\usepackage{booktabs} 

\usepackage{hyperref}

\usepackage[accepted]{icml2022}

\usepackage{amsmath}
\usepackage{amssymb}
\usepackage{mathtools}
\usepackage{amsthm}

\usepackage[capitalize,noabbrev]{cleveref}

\theoremstyle{plain}

\theoremstyle{definition}

\theoremstyle{remark}

\usepackage[disable,textsize=tiny]{todonotes}

\icmltitlerunning{Generating Synthetic Population}

\begin{document}

\twocolumn[
\icmltitle{Generating Synthetic Population}



\icmlsetsymbol{equal}{*}

\begin{icmlauthorlist}
\icmlauthor{Bhavesh Neekhra}{yyy}
\icmlauthor{Kshitij Kapoor}{yyy}
\icmlauthor{Debayan Gupta}{yyy}

\end{icmlauthorlist}

\icmlaffiliation{yyy}{Department of Computer Science, Ashoka University, India}

\icmlcorrespondingauthor{Bhavesh Neekhra}{bhavesh.neekhra\_phd18@ashoka.edu.in}

\icmlkeywords{Machine Learning, synthetic population, Generative Adversarial Networks (GAN), Iterative Proportional Fitting (IPF), Iterative Proportional Updating (IPU)}

\vskip 0.3in
]



\printAffiliationsAndNotice{}  

\begin{abstract}
In this paper, we provide a method to generate synthetic population at various administrative levels for a country like India. This synthetic population is created using machine learning and statistical methods applied to survey data such as Census of India 2011, IHDS-II, NSS-68th round, GPW etc. The synthetic population defines individuals in the population with characteristics such as age, gender, height, weight, home and work location, household structure, preexisting health conditions, socio-economical status, and employment. We used the proposed method to generate the synthetic population for various districts of India. We also compare this synthetic population with source data using various metrics. The experiment results show that the synthetic data can realistically simulate the population for various districts of India. 
\end{abstract}

\subsubsection*{Introduction}
\label{submission}
The Census Organisation in India has been publishing the tabulated results of Census since the initiation of modern Census in 1872 in various book forms.  As the information given by any respondent is treated as confidential, this micro-data is not publicly available.~\cite{isi-census} However Agent-based models, such as infectious disease modelling for Covid-19 require individual level information~\cite{Bonabeau02} to model and simulate the behavior of the system's constituent units (the agents) and their interactions. Though, within the 12th Five Year Plan (2012\-17) Indian Census set-up 18 Workstations country-wide for research on micro-data from Census, it lacked the flexibility which researchers  need in using such huge datasets. Thus it is crucial to have the ability to model population data and generate synthetic population at various administrative levels like, the Country, State, District, Sub-District, Town, Village or Ward in Town as the case may be. 

In this paper, we use a variety of data sources to generate a population of individuals and households with demographic attributes that are statistically identical to real data. This population is generated using a hybrid~\footnote{Our model is open-sourced at \url{https://github.com/bhaveshneekhra/synthpop} } of statistical methods and machine learning algorithms that are flexible enough to generate data at various administrative levels, ranging from small communities to states. The primary sources of data for these algorithms include the Census of India~\cite{census-2011}, the India Human Development Survey (IHDS)~\cite{Desai18}, the National Sample Survey (NSS)~\cite{nss18}, and the Gridded Population of the World (GPW)~\cite{ciesin2016}.

While the synthetic population should faithfully reproduce demographic statistics, it must also incorporate other realistic network structures, such as those appropriate to households and workplaces.~\footnote{Otherwise, we could end up, for example, with ``families'' composed entirely of toddlers, or workplaces with strange mixes of professions.} Because different kinds of data respond well to different techniques, a hybrid process is used to scale up these datasets. First, the data is cleaned to remove obvious inconsistencies. Next, a customized hybrid of Iterative Proportional Fitting (IPF)~\cite{beckman-1996, Deming1940}, Iterative Proportional Updating (IPU)~\cite{ipu2009}, and a specialized variant of a neural network, called Conditional-Tabular Generative Adversarial Network (CTGAN)~\cite{xu2019}, is used to generate new data. 

Briefly, Iterative Proportional Fitting finds a joint distribution that matches the marginals, while trying to stay as close to the sample distribution as possible. Iterative Proportional Updating is a heuristic iterative approach which can simultaneously match or fit to multiple distributions (constraints). Finally, Conditional-Tabular Generative Adversarial Networks is a method to model the tabular data distribution and sample rows from the distribution. A Generative Adversarial Network (GAN)~\cite{goodfellow14} uses two ``competing'' neural networks, the generator and the discriminator. The generator creates realistic samples with the goal that the discriminator should be unable to differentiate between a real sample and a generated sample. In this zero-sum game, capabilities of both the networks are enhanced iteratively. Critically, our techniques are designed to work seamlessly across data-scarce and data-rich areas; even if a particular area has error-prone or missing data, a synthetic population can still be generated, albeit of a lower quality.

\subsubsection*{The Population Generation Process}
    
We use IPF to generate a base population, using census data for the demographics and the IHDS survey dataset for personal and household attributes. The base population  thus consists of individual data and household data. We assign each household to an administrative unit within a district. 

We also experimented with CTGAN to generate a base population. The major advantage of IPU over CTGAN is that IPU is capable of matching individual level and household level characteristics of an individual while making sure that members of the household have a realistic age and gender joint distribution.

To assign job labels to individuals, the relevant data from the IHDS dataset is used. For the time-being, we classify individuals below the age of 18 as students, but could easily relax this assumption. A subset of the population is also assumed be home-bound. This subset consists of unemployed individuals, homemakers, infants and children under the age of 3 and elderly people over the retirement age. We use data from the NSS survey to determine the percentage of adult males and females in a city who are home-bound. A random independent sample is drawn from a Bernoulli distribution with this gender-based marginal value as a parameter in order to decide if an individual will be home-bound or not.


Each student in the population is assigned a school. Similarly, each working individual is assigned a workplace based upon their job label. We generate a synthetic latitude and longitude pair for each home, school and workplace in our dataset using GADM grid population density data~\cite{j-hijmans-2018}. We select a subset of grid points that lie within a given geographical boundary and sample grid-points with replacement grid points from the subset, weighing each point by the population density in the associated grid. We add independent random noise drawn from a uniform distribution to the latitudes and longitudes, rejecting those samples which fall outside the given geographical boundary. We follow this process to generate synthetic geolocation data for households, schools and workplaces.

To assign an individual a school, we sample from the list of schools within that geographical boundary, weighing each school by the inverse of the euclidean distance between it and the individual's home. This weighting factor increases the probability of assigning an individual a school that is closer to their home~\cite{rte-2019}. We follow a similar method to assign workplaces to adults. Additionally, based on every individual's job label, a workplace is assigned at random from a suitable subset of allowed workplaces.


A number of additional attributes are included in our synthetic population, including whether an individual uses public transport or whether an individual is an essential worker. These values are assigned using the individual's job label.

\begin{figure}[htb]
\vskip 0.2in
\begin{center}

\centerline{\includegraphics[width=\columnwidth]{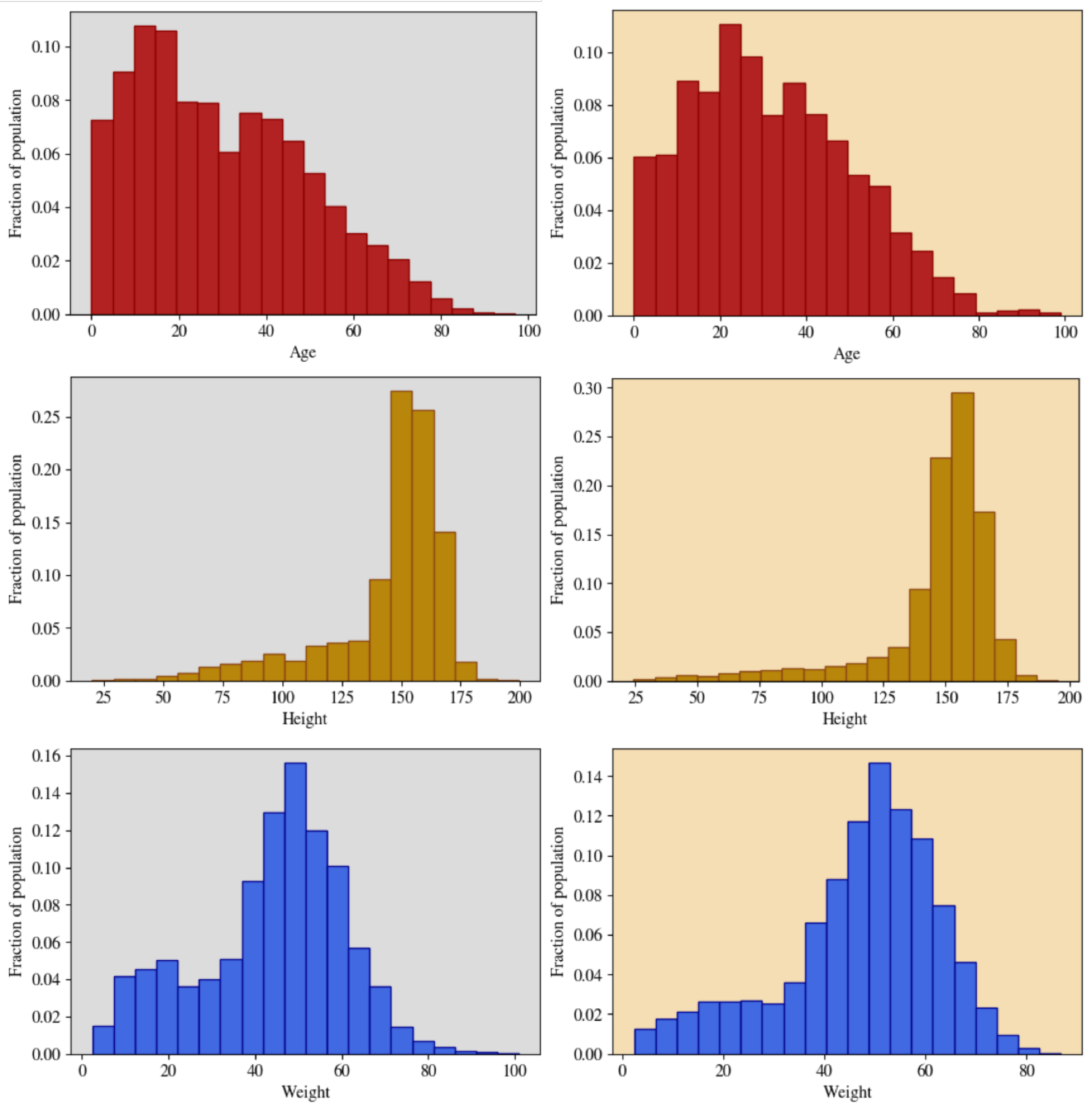}}
\caption{Histogram: Comparing source population (left) with synthetic population for the city of Mumbai in India}
\label{hist-plot-comparison}
\end{center}
\vskip -0.2in
\end{figure}

\begin{figure}[htb]
\vskip 0.2in
\begin{center}

\centerline{\includegraphics[width=\columnwidth]{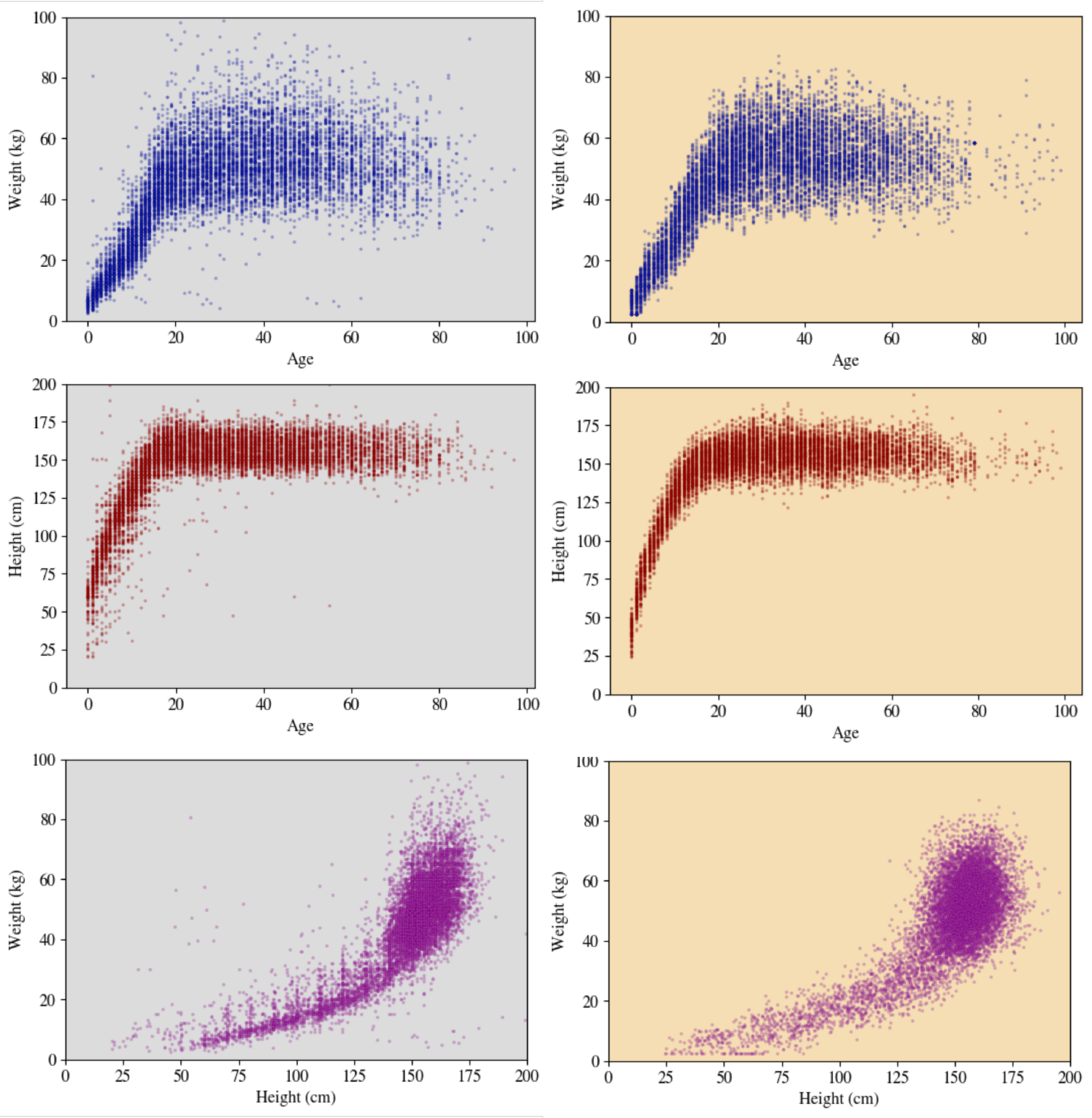}}
\caption{Scatter plot: Comparing source population (left) with synthetic population for the city of Mumbai in India}
\label{scatter-plot-comparison}
\end{center}
\vskip -0.2in
\end{figure}

\subsubsection*{Population Verification Metrics}
\begin{figure}[H]
\vskip 0.2in
\begin{center}
\centerline{\includegraphics[width=\columnwidth]{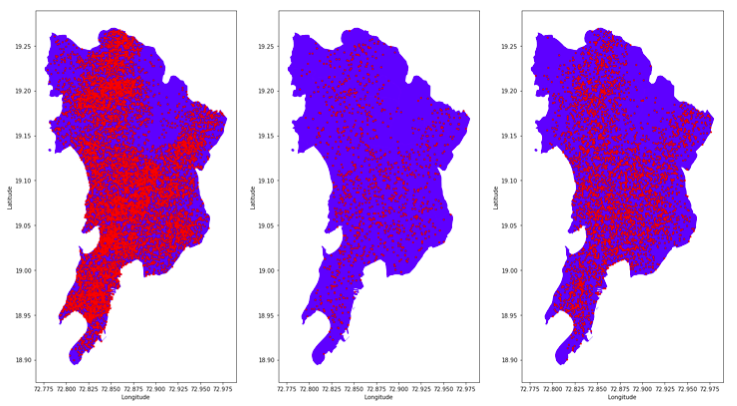}}
\caption{Geographical distribution of Households, Schools and Workplaces, respectively, in the Synthetic Population}
\label{geo-plots}
\end{center}
\vskip -0.2in
\end{figure}

To compare and verify the the generated synthetic population with the survey data, we used several methods. We used the Bhattacharya
distance to quantify the similarity of the joint age-height and age-weight distributions. In addition, apart from comparing the two
populations visually as seen in \cref{hist-plot-comparison} and \cref{scatter-plot-comparison}, we have also used a number of
other metrics such as statistical likelihood techniques (CS-test, KS-test). We also visualise the geographical spread of the households, schools and the workplaces in the population as in \cref{geo-plots}. The visual comparison shows the synthetic population resembles the real population.

\subsubsection*{Conclusion}
Lack of data due to access and privacy issues results in poor model design. To tackle this issue, we propose a hybrid method to generate a synthetic population. We also provide a combination of metrics to verify the generated data. In ongoing work, we are generating the synthetic population for the entire country of India. As future work, we want to explore the possibility of modelling more nuanced and complex features for the synthetic population.

\section*{Acknowledgements}
The authors are grateful for support from the Mphasis F1 Foundation and the Bill and Melinda Gates Foundation, Grant No: R/BMG/PHY/GMN/20.

\bibliography{my_paper}
\bibliographystyle{icml2022}

\newpage
\appendix
\onecolumn
\section{Sample Synthetic Population}
A sample of ten individuals in the synthetic population is reproduced below. The columns are split to fit in the page.

\begin{table}[!htb]
		\begin{tabular}{c|c|c|c|c|c|c|c|c}
			Age & SexLabel & Height & Weight & HHID & H\_Lat & H\_Lon & District & AdminUnitName \\
			\hline
			56 & Male & 165.970 & 44.381 & 5.190e+10 & 19.024 & 72.911 & Mumbai & M/E \\
			10 & Male & 138.150 & 34.388 & 5.190e+10 & 19.074 & 72.832 & Mumbai & H/W \\
			3 & Female & 65.460 & 7.205 & 1.038e+11 & 19.258 & 72.867 & Mumbai & R/N \\
			63 & Male & 170.400 & 59.962 & 1.038e+11 & 19.046 & 72.933 & Mumbai & M/E \\
			37 & Female & 143.530 & 60.381 & 5.190e+10 & 19.189 & 72.837 & Mumbai & P/N \\
			46 & Male & 163 & 63.064 & 1.038e+11 & 19.071 & 72.889 & Mumbai & L \\
			35 & Male & 165.660 & 70.111 & 1.038e+11 & 19.073 & 72.906 & Mumbai & N \\
			7 & Female & 119.680 & 21.093 & 1.038e+11 & 19.013 & 72.886 & Mumbai & M/W \\
			29 & Male & 152.520 & 59.198 & 5.190e+10 & 19.131 & 72.898 & Mumbai & S \\
			74 & Female & 169.560 & 37.874 & 5.190e+10 & 18.952 & 72.797 & Mumbai & D \\
		\end{tabular} 
         \end{table}
\begin{table}[!htb]   
		\begin{tabular}{c|c|c|c|c|c|c}
			AdminUnitLatitude & AdminUnitLongitude & Religion & Caste & JobLabel & JobID & WorkPlaceID \\
			\hline
			19.056 & 72.922 & Hindu & other & Carpenters & 81 & 2.001e+12 \\
			19.056 & 72.835 & Hindu & other & Student & 199 & 0 \\
			19.120 & 72.852 & Hindu & other & Homebound & 0 & 0 \\
			19.056 & 72.922 & Hindu & other & Homebound & 0 & 0 \\
			19.188 & 72.842 & Hindu & other & Homebound & 0 & 0 \\
			19.070 & 72.879 & buddhist & other & Labour nec & 99 & 2.001e+12 \\
			19.084 & 72.906 & Hindu & SC & Construction & 95 & 2.001e+12 \\
			19.061 & 72.899 & Hindu & other & Student & 199 & 0 \\
			19.139 & 72.930 & Hindu & other & Construction & 95 & 2.001e+12 \\
			18.963 & 72.813 & Hindu & other & Construction & 95 & 2.001e+12 \\
		\end{tabular} 
    \end{table}
\begin{table}[!htb]        
  
		\begin{tabular}{c|c|c|c|c|c}
			W\_Lat & W\_Lon & essential\_worker & Adherence\_to\_Intervention & PublicTransport\_Jobs & school\_id \\
			\hline
			19.036 & 72.867 & 0 & 0.900 & 1 & 0 \\
			nan & nan & 0 & 0.800 & 1 & 2.001e+12 \\
			nan & nan & 0 & 1 & 1 & 0 \\
			nan & nan & 0 & 1 & 1 & 0 \\
			nan & nan & 0 & 0 & 1 & 0 \\
			19.071 & 72.890 & 0 & 0.900 & 1 & 0 \\
			19.192 & 72.865 & 0 & 0 & 1 & 0 \\
			nan & nan & 0 & 1 & 1 & 2.001e+12 \\
			19.155 & 72.883 & 0 & 0.200 & 1 & 0 \\
			19.024 & 72.850 & 0 & 1 & 1 & 0 \\
		\end{tabular}
  \end{table}
\begin{table}[!htb]
		\begin{tabular}{c|c|c|c|c|c|c}
			school\_lat & school\_long & public\_place\_id & public\_place\_lat & public\_place\_long & Agent\_ID & PSUID \\
			\hline
			nan & nan & 3.001e+12 & 19.024 & 72.910 & 5.191e+10 & 1 \\
			19.177 & 72.867 & 3.001e+12 & 19.179 & 72.828 & 5.191e+10 & 20 \\
			nan & nan & 3.001e+12 & 19.098 & 72.844 & 5.191e+10 & 9 \\
			nan & nan & 3.001e+12 & 19.148 & 72.842 & 5.190e+10 & 14 \\
			nan & nan & 3.001e+12 & 19.189 & 72.830 & 5.190e+10 & 17 \\
			nan & nan & 3.001e+12 & 19.092 & 72.889 & 5.191e+10 & 5 \\
			nan & nan & 3.001e+12 & 19.059 & 72.895 & 5.190e+10 & 8 \\
			18.962 & 72.838 & 3.001e+12 & 19.223 & 72.866 & 5.191e+10 & 22 \\
			nan & nan & 3.001e+12 & 19.129 & 72.913 & 5.191e+10 & 13 \\
			nan & nan & 3.001e+12 & 19.046 & 72.851 & 5.190e+10 & 1 \\
		\end{tabular}
  \end{table}
  \begin{table}[]

	\begin{tabular}{c|c|c|c|c|c|c}
			M\_Fever & M\_Diarrhea & M\_Cataract & M\_Heart\_disease & M\_Diabetes & M\_Leprosy & M\_Cancer \\
			\hline
			0 & 0 & 0 & 0 & 0 & 0 & 0 \\
			0 & 0 & 0 & 0 & 0 & 0 & 0 \\
			0 & 0 & 0 & 0 & 0 & 0 & 0 \\
			0 & 0 & 0 & 0 & 0 & 0 & 0 \\
			0 & 0 & 0 & 0 & 0 & 0 & 0 \\
			0 & 0 & 0 & 0 & 0 & 0 & 0 \\
			0 & 0 & 0 & 0 & 0 & 0 & 0 \\
			0 & 0 & 0 & 0 & 0 & 0 & 0 \\
			0 & 0 & 0 & 0 & 0 & 0 & 0 \\
			0 & 0 & 0 & 0 & 0 & 0 & 0 \\
		\end{tabular}

  \begin{tabular}{c|c|c}
			M\_Asthma & M\_Paralysis & M\_Epilepsy \\
			\hline
			0 & 0 & 0 \\
			0 & 0 & 0 \\
			0 & 0 & 0 \\
			0 & 0 & 0 \\
			0 & 0 & 0 \\
			0 & 0 & 0 \\
			0 & 0 & 1 \\
			0 & 0 & 1 \\
			0 & 0 & 1 \\
			0 & 0 & 0 \\
		\end{tabular}
\end{table}


\end{document}